\documentclass[conference]{IEEEtran}
\IEEEoverridecommandlockouts
\usepackage{cite}
\usepackage{graphicx}  
\usepackage{amsmath,amssymb,amsfonts}
\usepackage{algorithm}
\usepackage{textcomp}
\usepackage{xcolor}
\usepackage{algpseudocode}
\usepackage{booktabs}       
\usepackage{makecell}       
\usepackage[table]{xcolor}  
\usepackage{multirow}

\def\BibTeX{{\rm B\kern-.05em{\sc i\kern-.025em b}\kern-.08em
    T\kern-.1667em\lower.7ex\hbox{E}\kern-.125emX}}
\begin{document}

\title{\huge VeRA+: Vector-Based Lightweight Digital Compensation for Drift-Resilient RRAM In-Memory Computing}
 
\author{
Weirong Dong$^{*}$$^{\dagger}$, Kai Zhou$^{*}$, Zhen Kong$^{*}$, Zhengke Yang$^{*}$, Quan Cheng$^{\dagger}$, Haoyuan Li$^{\dagger}$,  \\ Junkai Huang$^{*}$, Jun Lan$^{*}$, Yida Li$^{*}$, Masanori Hashimoto$^{\dagger}$, Longyang Lin$^{*}$ \\
\IEEEauthorblockA{
    $^*$School of Microelectronics, Southern University of Science and Technology, Shenzhen, China \\
    $^{\dagger}$Department of Informatics, Kyoto University, Kyoto, Japan \\
    \thanks{Accepted at the 2026 Design Automation Conference (DAC).}
}
}

\maketitle

\begin{abstract}

RRAM-based in-memory computing (IMC) offers high energy efficiency
but suffers from conductance drift that severely degrades long-term
accuracy. Existing approaches including retraining, noise-aware
training, and Batch Normalization (BN)-based calibration either require RRAM rewriting,
demand large storage overhead, or rely on online correction. We
propose VeRA+, a lightweight drift compensation framework that
reuses shared projection matrices and introduces only two compact
drift-specific vectors per drift level. A drift-aware scheduling
algorithm offline-trains a small set of VeRA+ parameters and
selects the appropriate set over time without any on-chip
retraining or data replay. VeRA+ preserves up to 99.77\% of the drift-free accuracy after ten years of simulated drift and reduces storage overhead by more than three orders of magnitude compared with BN-based calibration. To validate VeRA+ under realistic device behavior, we extract one-week drift statistics from measurements on our fabricated 1T1R RRAM devices and use them to simulate realistic drifted weights. Under these measured drift conditions,
VeRA+ achieves accuracy close to the drift-free baseline, providing an
efficient and practical solution for long-term drift resilience in RRAM-IMC.
\end{abstract}



\section{Introduction}
The growing demand for efficient machine learning at the edge has spurred intense interest in resistive random-access memory (RRAM)-based in-memory computing (IMC). By co-locating data storage and computation in dense non-volatile arrays, RRAM-based IMC circumvents the von Neumann bottleneck and achieves remarkable energy and area efficiency \cite{RIMC1,RIMC2,RIMC3}. Despite these advantages, RRAM-IMC faces a fundamental obstacle to long-term deployment: conductance drift \cite{relaxation2}. Unlike static programming errors \cite{programming_noise}, which occur during weight initialization,
conductance drift refers to the gradual, time-dependent degradation of programmed conductance values due to intrinsic physical relaxation processes in the RRAM cells \cite{relaxation3}.  
Consequently, even accurately programmed weights can deviate significantly over hours, days, or years of operation.
For example, our experiment described later shows that a ResNet-20 model that achieves 68\% accuracy in CIFAR-100 dataset without drift drops to 19\% after one month of simulated drift, 
with performance continuing to degrade under longer-term scenarios. This underscores the urgency of developing effective drift compensation mechanisms.

The most prominent prior approach is variation-aware training to mitigate programming noise, combined with adaptive Batch Normalization (BN) to compensate for long-term drift~\cite{variationaware}. However, this method requires storing about 5\% of the training dataset on-chip for chip-in-the-loop BN statistics computation, which imposes considerable storage overhead. On-chip retraining \cite{onchip, onchip2}, \sloppy though conceptually appealing, is throttled by RRAM’s slow write-and-verify cycle \cite{write_and_verify}, limited endurance \cite{endurance} and large training dataset. 


\begin{figure}[t]
    \centering
    \includegraphics[width=0.5\textwidth, keepaspectratio]{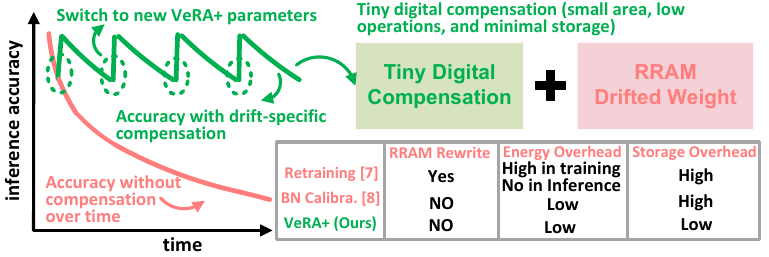}
    \vspace{-22pt}
    \caption{Accuracy loss from RRAM conductance drift and the benefit of lightweight drift-specific digital compensation.
    }\vspace{-22pt}
    \label{temporal_compensation}

\end{figure}

To achieve lifecycle-aware drift compensation without such prohibitive cost, we propose Vector-based Random Matrix Adaptation plus (VeRA+), a lightweight digital post-deployment calibration framework tailored with hybrid SRAM–RRAM IMC systems, as illustrated in Fig.~\ref{temporal_compensation}. VeRA+ builds upon the principle of parameter-efficient adaptation but is redesigned for convolutional architectures and long-term reliability management. Specifically, VeRA+ shares projection matrices across layers and drift levels, introducing only two compact drift-specific scaling vectors stored in SRAM.

Our key contributions are as follows:
\begin{itemize}
    \item We propose VeRA+, a long-term drift-aware compensation framework for hybrid SRAM–RRAM IMC. VeRA+ leverages parameter-efficient adaptation techniques to correct drift without RRAM rewriting or online retraining.
    
    \item We introduce VeRA+, which improves upon VeRA to better support CNNs and further reduces the storage, area and energy overhead across different drift levels.

    \item We develop a drift-aware scheduling algorithm that automatically discovers when drift significantly impacts accuracy and allocates new compensation vectors only when necessary, enabling reliable long-term operation with minimal storage cost.

    \item We validate VeRA+ across diverse neural architectures \sloppy (ResNet20/32/50 and BERT base/large) with both an existing drift model and our measured drift data from our fabricated RRAM devices, demonstrating its scalability, generality, and robustness for long-term reliability management of RRAM-IMC systems.
\end{itemize}

\section{Background}
\subsection{Conductance Drift Model}
\label{subsec:driftmodel}
To assess and mitigate the impact of drift on computation accuracy, several modeling approaches have been reported \cite{relaxation2, other2}.
Among these, the experiments in this work adopt the statistical
drift model from IBM’s Analog AI Hardware Kit~\cite{IBM} because it provides well-calibrated, device-verified model on long-term RRAM drift \cite{drift_model}. The drifted conductance $g_\text{real}(t)$ for the target conductance $g_\text{target}$ is modeled as follows.
\begin{equation}
g_\text{drift}(t) \sim \mathcal{N}(\mu_\text{drift}(t), \sigma^2_\text{drift}(t)), \label{eq1}
\end{equation}
\begin{equation}
\mu_\text{drift}(t) = 0.089 \cdot \log(t)\ [\mu\text{S}], \label{eq2}
\end{equation}
\begin{equation}
\sigma_\text{drift}(t) = 0.042 \cdot \log(t) + 0.4118\ [\mu\text{S}]. \label{eq3}
\end{equation}
where $t$ is the elapsed time in seconds. To additionally account for device-to-device drift variations, we apply a multiplicative Gaussian perturbation to each device instance:
\begin{equation}
g_\text{real}(t) = (g_\text{target} + g_\text{drift}(t)) \cdot (1 + \epsilon),\quad \epsilon \sim \mathcal{N}(0,0.05^2), \label{eq4}
\end{equation}
which introduces approximately $\pm5\%$ variation across devices.

\begin{table}[t]
  \caption{RRAM and SRAM IMC Hardware at 22\,nm node}
  \vspace{-4pt}
  \label{tab:hardware}
  \centering
  \begin{tabular}{lll}
    \toprule
    \textbf{Metric} & \textbf{RRAM-IMC} & \textbf{SRAM-IMC} \\
    \midrule
    Energy Efficiency$^{*}$ & 209\,TOPS/W & 89\,TOPS/W \\
    Memory Density & 2.53\,Mb/mm$^2$ & 0.31\,Mb/mm$^2$ \\
    Volatility & Non-volatile & Volatile \\
    \bottomrule
  \end{tabular}
  \begin{minipage}{0.95\linewidth}
    \footnotesize\emph{Note:} Energy efficiency is measured under the int4 setting.
  \end{minipage}
  \vspace{-18pt}
\end{table}

\subsection{RRAM-IMC vs SRAM-IMC}
A key question is whether digital computation can effectively compensate analog errors in RRAM-IMC without undermining its efficiency. Table~\ref{tab:hardware} compares RRAM-IMC~\cite{rram_imc} and SRAM-IMC~\cite{sram_2021} at the same 22-nm node. RRAM-IMC offers superior density and energy efficiency, while SRAM-IMC provides higher precision but with nearly 8× lower memory density and over 2× lower energy efficiency. These results emphasize that drift compensation must incur minimal parameter overhead to keep the SRAM-based digital module compact and preserve the system-level advantage of RRAM-IMC.

\subsection{Parameter-Efficient Compensation}
Parameter-efficient fine-tuning (PEFT) has emerged as a promising approach for lightweight model adaptation. Among these methods, Low-Rank Adaptation (LoRA)~\cite{LoRA} is widely used due to its simplicity and effectiveness. LoRA injects trainable low-rank matrices $(A, B)$ into frozen backbones $(W_0)$ to obtain a fine-tuned weight,
\begin{equation}
W_{\mathrm{eff}}X = W_0X + BAX, \label{eq:lora}
\end{equation}
where $X$ denotes the input vector with $A \in \mathbb{R}^{r \times d}$ and $B \in \mathbb{R}^{d \times r}$, and a low rank $r \ll d$. By updating only the small low-rank components $A$ and $B$, LoRA greatly reduces the number of trainable parameters.  

VeRA~\cite{vera} further improves efficiency by replacing layer-specific $A$ and $B$ with shared random projection matrices $(A_\mathrm{R}, B_\mathrm{R})$ and introducing two trainable scaling vectors for each layer and drift level: 
\begin{equation}
W_{\mathrm{eff}} X
= W_0X + b \odot \bigl( B_{\mathrm{R}} \bigl( d \odot (A_{\mathrm{R}} X) \bigr) \bigr),
\end{equation}

where $A_\mathrm{R} \in \mathbb{R}^{r \times d}$ and $B_\mathrm{R} \in \mathbb{R}^{d \times r}$ are shared random projections, while $d \in \mathbb{R}^{r}$ and $b \in \mathbb{R}^{d}$ are lightweight scaling vectors, $\odot $ denotes the Hadamard product. This formulation directly modulates the output activations $X$ through low-rank adaptation, drastically reducing parameter overhead while maintaining accuracy.

\subsection{Prior Drift-Compensation Approaches}

variation-aware training \cite{dac} injects conductance drift and device variation into the
training process to improve robustness. However, the magnitude and
statistics of drift change substantially over time, making it difficult for a
single offline-trained model to remain robust across the entire drift horizon.
A model trained to tolerate short-term drift (seconds to hours) typically fails
under long-term drift, where the accumulated deviation becomes
significantly larger. Conversely, a model trained for long-term drift (months
to years) cannot effectively handle the much
smaller perturbations present during early deployment. 

A more widely adopted approach is BN-based post-training calibration~\cite{variationaware},
which recomputes BN statistics from a subset of the training data to
counteract drift. Although this avoids rewriting the RRAM arrays, it demands
storing a non-negligible portion of the training set and periodically updating
BN statistics. Moreover, modern inference pipelines fold BN into convolution
weights; thus, BN-based calibration prevents operator fusion and introduces
additional computational overhead.

These limitations motivate compensation techniques that avoid retraining,
data replay, and unfolded-BN inference while maintaining long-term drift
robustness.

\begin{figure}[t]
    \centering
    \includegraphics[width=0.5\textwidth, keepaspectratio]{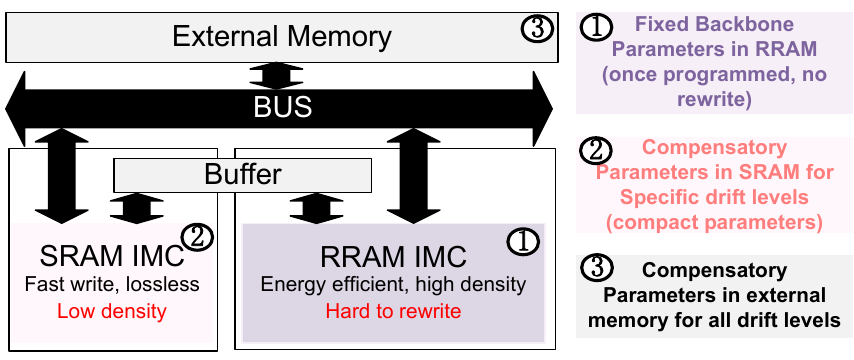}
    \vspace{-18pt}
    \caption{Hybrid deployment for drift-aware compensation. 
}\vspace{-3pt}
    \label{fig2}
    \vspace{-13pt} 
\end{figure}

\section{Proposed Method}

This section introduces VeRA+ in four parts. 
Section~III-A outlines the basic idea of using compact scaling vectors for drift compensation. 
Section~III-B presents the hybrid SRAM–RRAM deployment. 
Section~III-C details CNN-specific designs such as cross-layer sharing and the 1×1 kernel scheme. 
Section~III-D describes the drift-aware scheduling and training that minimize the number of compensation branches.

\subsection{Basic Idea}

Our baseline idea is to compute $W_0X$ in RRAM-IMC and $BAX$ in SRAM-IMC, where the low-rank matrices $A$ and $B$ compensate for drift-induced changes in $W_0$. Since the RRAM backbone remains fixed after programming, only the low-rank parameters need to be updated. When conductance drift changes $W_0$ to $W(t)$, the compensation must satisfy
\begin{equation}
(W(t)-W(0))X \simeq B(t)A(t)X.
\end{equation}

However, storing a separate pair of matrices $A(t)$ and $B(t)$ for every layer and every drift level is prohibitive, even when drift is coarsely quantized. The storage cost grows quickly with both network depth and the number of drift levels. Therefore, an effective drift-aware compensation scheme must minimize not only the SRAM-side computation but also the cumulative parameter cost over time.

VeRA addresses this issue by reusing fixed projection matrices $A_{\mathrm R}$ and $B_{\mathrm R}$ across all layers, replacing $A(t)$ and $B(t)$ with two small drift-dependent vectors $b(t)$ and $d(t)$. To further reduce overhead, VeRA+ shares $A_{\mathrm R}$ and $B_{\mathrm R}$ across both layers and drift levels:
\begin{equation}\label{eq:vera}
W(t)-W(0) \simeq \operatorname b(t) \odot \bigl( B_{\mathrm{R}} \bigl( d(t) \odot (A_{\mathrm{R}} X) \bigr) \bigr),
\end{equation}
where $b(t)$ and $d(t)$ are the only drift-specific parameters.

Since $b(t)$ and $d(t)$ vary continuously with time, directly storing them is impractical. We instead approximate them using a small set of discrete vectors:
\begin{equation}\label{eq:discrete}
b(t) \simeq b_k,\quad d(t) \simeq d_k,\quad t\in[t_k,t_{k+1}],
\end{equation}
where each pair $(b_k,d_k)$ corresponds to one quantized drift level.
This reduces storage from two matrices per level to two vectors, making drift-aware compensation lightweight and hardware-friendly.

\subsection{Drift-Aware RRAM- and SRAM-IMC Hybrid Architecture}

Fig.~\ref{fig2} shows the hardware architecture assumed in this work. The architecture consists of three memory systems:
\begin{itemize}
\item \textbf{RRAM-IMC}: Stores the backbone weights $W_0$ at time 0 and performs computations of the backbone neural network. The conductance values vary over time, but no rewrite is performed after deployment. 
\item \textbf{SRAM-IMC}: Stores the currently active  $(b_k, d_k)$ and the shared projection matrices $A_R$ and $B_R$, and computes the compensation of Eq.(\ref{eq:vera}).
\item \textbf{External Memory}: The complete set of $b_k$ and $d_k$, as well as $A_R$ and $B_R$, are stored in ROM/Flash and loaded into SRAM-IMC when needed.
\end{itemize}

For deploying this idea, two issues must be further addressed.
(1) A network consists of different types and sizes of computations, and the digital compensation hardware must accommodate them efficiently.
(2) The number of vector sets ($b_k$, $d_k$) should be minimized while maintaining network accuracy within an acceptable range.

\subsection{CNN-Specific Compensation and Cross-Layer Sharing}
LoRA and VeRA were originally developed for transformer models and are not directly tailored to convolutional architectures. Convolutional networks introduce two specific challenges: (i) input and output channel dimensions vary across depth, and (ii) spatial weight sharing already provides parameter efficiency, so a naive application of LoRA or VeRA leads to unnecessary overhead.

To address (i), VeRA+ instantiates large global matrices $A_{max} \in \mathbb{R}^{r \times d_{\max}^{\mathrm{in}}}$ and $B_{max} \in \mathbb{R}^{d_{\max}^{\mathrm{out}} \times r}$ for the entire network. For each layer $i$, submatrices are selected as
\[
A_R^{(i)} = A_{\max}[:,\, 1:d^{(i)}_{\mathrm{in}}], \quad
B_R^{(i)} = B_{\max}[1:d^{(i)}_{\mathrm{out}}, :]
\]
which means that the first $d^{(i)}_{\mathrm{in}}$ columns of $A_{\max}$ 
and the first $d^{(i)}_{\mathrm{out}}$ rows of $B_{\max}$ are used for 
layer $i$. This strategy allows each layer to use a customized slice of 
the global matrices while still keeping $A_R$ and $B_R$ shared across all 
drift levels. 

\vspace{-4pt}
\begin{algorithm}[!htbp]
\caption{VeRA+: Drift-Aware Scheduling and Training}
\label{alg:star}
\begin{algorithmic}[1]
\Require Quantized weights $W_{\text{RRAM}}$, fixed $A_R$, $B_R$; drift model $\mathcal{M}_{\text{drift}}(t)$; accuracy threshold $a_{\text{thr}}$; dataset $\mathcal{D}$; max time $T_{\max}$ 
\Ensure Drift points $\mathcal{T}$ and trained branches $\{(b(t), d(t))\}$
\State $t \gets 1$;\; $\mathcal{T} \gets \{t\}$;\; 
\While{$t < T_{\max}$}
    \State $t \gets 1.5 \times t$
    \State $(\mu(t), \sigma(t)) \gets \text{EVALSTATS}(t,  \mathcal{M}_{\text{drift}})$ \Comment{mean/std over drift samples}
    \If{$\mu - 3\sigma < a_{\text{thr}}$} \Comment{lower confidence bound below floor}
        \State $\mathcal{T} \gets \mathcal{T} \cup \{t\}$;\; Initialize $b(t), d(t)$
        \For{mini-batch $(x, label)$ in $\mathcal{D}$}
            
            \State $W_{\text{drift}} \gets \textsc{DriftInject}(W_{\text{RRAM}}, t)$
            \Comment{a new drift instance is sampled for each mini-batch}
            \State $\hat{y} \gets  W_{drift} X + W_{\mathrm{VeRA}} X$
            \State $\mathcal{L} \gets \textsc{Loss}(\hat{y}, label)$
            \State \textsc{Update}$(b(t), d(t);\,\nabla \mathcal{L})$ 
        \EndFor
    \EndIf
\EndWhile
\State \Return $\mathcal{T}$ and $(b(t), d(t))$
\end{algorithmic}
\end{algorithm}
\vspace{-4pt}
To address (ii), VeRA+ adopts an aggressive lightweight compensation scheme. Unlike official LoRA/VeRA for CNN, which uses low-rank matrices $A$ with shape $[r \times K,\ C_{\text{in}} \times K]$ and $B$ with shape $[C_{\text{out}} \times K,\ r \times K]$, where $K$ denotes the kernel size. VeRA+ generates compensation weights in a $1 \times 1$ form with $A_R$ of shape $[r,\ C_{\text{in}}]$ and $B_R$ of shape $[C_{\text{out}},\ r]$. As a result, both parameter and compute overhead are significantly reduced by up to $9\times$ for $3 \times 3$ kernels while maintaining full convolutional functionality.

\subsection{Training with Minimum Number of VeRA+ Sets}
We first train the backbone weights $W_0$ using quantization-aware training \cite{QAT} to obtain a quantized model, 
which will be then programmed into RRAM. We subsequently perform drift-inject training on top of this backbone. 

 To minimize the number VeRA+ compensation sets, we devise Algorithm~\ref{alg:star}. We first describe the inner procedure that trains scaling vectors under simulated drift at time $t$ (Lines 7 to 12). We then explain the overall behavior of the algorithm.

\subsubsection{Training under Simulated Drift at Time $t$}
For a given drift time $t$, we train on the full dataset divided into mini-batches.  
In each mini-batch, drift is injected once by sampling from the log-normal statistical model in Eqs.~(\ref{eq1})–(\ref{eq4}) to generate a perturbed weight instance $W_{\text{drift}}$.   We temporarily inject drift into the frozen backbone weights and perform the forward pass with the drifted weights. Before the backward pass, the original weights are restored so that only the drift-specific scaling vectors $b(t)$ and $d(t)$ are updated while the backbone and shared projection matrices $A_R$, $B_R$ remain frozen. 
Each mini-batch applies a newly sampled drift instance, allowing the compensation vectors to learn robustness across many hardware realizations, yielding compact compensation parameters $\{b(t), d(t)\}$ that restore accuracy under drift at time $t$.  All compensation sets are trained offline using this statistical-variation model, and deployment merely loads the pre-trained vectors into SRAM without any on-chip retraining. In our experiments, each drift level was trained for only 3 epochs with a mini-batch size of 64.

\subsubsection{Training across Multiple Drift Levels}

Algorithm~\ref{alg:star} advances time $t$ exponentially, since the conductance drift in Eqs.~(\ref{eq2})--(\ref{eq3}) is modeled as a log-normal function (Line 3). The multiplier of 1.5 is an empirical value used in our experiment in the next section, and it can be adjusted.

For each advanced drift time $t$, we evaluate the accuracy and compare it against the required threshold $a_{\mathrm{thr}}$ (Line 5). Since the achievable accuracy varies across instances of drifted weights, the EVALSTATS function in Line 4 estimates the mean $m(t)$ and standard deviation $\sigma(t)$ of the network accuracy by generating multiple drifted-weight instances using the statistical drift model.

A new set is trained (Lines 6 to 12) only when the lower bound of the 99.7\% confidence interval, $\mu(t) - 3\sigma(t)$, drops below a target accuracy threshold $a_{\mathrm{thr}}$. At this point, the
corresponding drift time $t$ and its trained VeRA+ parameters are recorded. During deployment, the chip
simply selects the appropriate pre-trained set according to the elapsed time provided by a lightweight
timer or host controller. The confidence interval can be adjusted depending on reliability requirements.

\begin{table*}[t]
  \caption{Model accuracy degradation over time and compensation results (mean $\pm$ std) with $r=1$}
  \vspace{-4pt}
  \label{tab:drift_results}
  \centering
  \begin{tabular}{p{1.4cm}ccccccccccc}

    \toprule
    \textbf{Model} & \textbf{Dataset} & \textbf{Drift Free} & \textbf{1s} & \textbf{1h} & \textbf{1d} & \textbf{1mon} & \textbf{1y} & \textbf{10y} & \textbf{1y comp.} & \textbf{10y comp.} \\
    \midrule
    \multirow{2}{*}{ResNet-20} 
      & CIFAR-10  & 92.20 & 92.09{\scriptsize$\pm$0.1} & 90.44{\scriptsize$\pm$0.2} & 87.27{\scriptsize$\pm$0.4} & 81.75{\scriptsize$\pm$0.7} & 75.64{\scriptsize$\pm$1.0} & \cellcolor{red!10}68.46{\scriptsize$\pm$1.5} & 92.03{\scriptsize$\pm$0.2} & 
      \cellcolor{green!10}92.01{\scriptsize$\pm$0.2} \\
      & CIFAR-100 & 68.84 & 68.65{\scriptsize$\pm$0.1} & 48.85{\scriptsize$\pm$0.7} & 35.21{\scriptsize$\pm$0.8} & 18.99{\scriptsize$\pm$0.8} & 12.85{\scriptsize$\pm$0.6} &  \cellcolor{red!10}9.09{\scriptsize$\pm$0.4} & 67.06{\scriptsize$\pm$0.2} & \cellcolor{green!10}66.91{\scriptsize$\pm$0.3} \\
    \midrule
    \multirow{2}{*}{ResNet-32} 
      & CIFAR-10  & 93.75 & 93.70{\scriptsize$\pm$0.1} & 91.17{\scriptsize$\pm$0.2} & 88.15{\scriptsize$\pm$0.3} & 82.19{\scriptsize$\pm$0.5} & 75.53{\scriptsize$\pm$0.9} & \cellcolor{red!10}67.45{\scriptsize$\pm$1.1} & 92.95{\scriptsize$\pm$0.1} & \cellcolor{green!10}92.90{\scriptsize$\pm$0.2} \\
      & CIFAR-100 & 71.46 & 71.12{\scriptsize$\pm$0.1} & 54.52{\scriptsize$\pm$0.6} & 37.51{\scriptsize$\pm$0.8} & 19.86{\scriptsize$\pm$0.7} & 11.70{\scriptsize$\pm$0.5} &  \cellcolor{red!10}7.04{\scriptsize$\pm$0.3} & 69.40{\scriptsize$\pm$0.2} & \cellcolor{green!10}69.21{\scriptsize$\pm$0.3} \\
    \midrule
    ResNet-50 
      & ImageNet-1K & 76.09 & 72.92{\scriptsize$\pm$0.2} & 33.45{\scriptsize$\pm$0.3} & 0.72{\scriptsize$\pm$0.1} & 0.15{\scriptsize$\pm$0.0} & 0.13{\scriptsize$\pm$0.0} & \cellcolor{red!10}0.11{\scriptsize$\pm$0.0} & 72.08{\scriptsize$\pm$0.2} & \cellcolor{green!10}70.33{\scriptsize$\pm$0.2} \\
    \midrule
    \multirow{2}{*}{BERT-base} 
      & QQP       & 90.10 & 89.93{\scriptsize$\pm$0.1} & 89.87{\scriptsize$\pm$0.1} & 89.55{\scriptsize$\pm$0.1} & 89.39{\scriptsize$\pm$0.1} & 89.13{\scriptsize$\pm$0.2} & \cellcolor{red!10}88.76{\scriptsize$\pm$0.3} & 90.06{\scriptsize$\pm$0.1} & \cellcolor{green!10}90.05{\scriptsize$\pm$0.1} \\
      & SST-5     & 53.98 & 53.42{\scriptsize$\pm$0.5} & 52.69{\scriptsize$\pm$0.7} & 52.22{\scriptsize$\pm$0.8} & 51.97{\scriptsize$\pm$0.9} & 51.23{\scriptsize$\pm$0.9} & \cellcolor{red!10}50.25{\scriptsize$\pm$1.2} & 53.54{\scriptsize$\pm$0.2} & \cellcolor{green!10}53.48{\scriptsize$\pm$0.2} \\
    \midrule
    \multirow{2}{*}{BERT-large} 
      & QQP       & 91.07 & 90.80{\scriptsize$\pm$0.1} & 90.42{\scriptsize$\pm$0.2} & 90.20{\scriptsize$\pm$0.2} & 89.78{\scriptsize$\pm$0.3} & 89.47{\scriptsize$\pm$0.3} & \cellcolor{red!10}89.01{\scriptsize$\pm$0.4} & 91.02{\scriptsize$\pm$0.1} & \cellcolor{green!10}91.01{\scriptsize$\pm$0.1} \\
      & SST-5     & 55.66 & 54.36{\scriptsize$\pm$0.6} & 53.28{\scriptsize$\pm$0.6} & 52.81{\scriptsize$\pm$0.9} & 51.94{\scriptsize$\pm$1.1} & 51.40{\scriptsize$\pm$1.3} & \cellcolor{red!10}50.45{\scriptsize$\pm$1.3} & 55.29{\scriptsize$\pm$0.2} & \cellcolor{green!10}55.12{\scriptsize$\pm$0.3} \\
    \bottomrule
  \end{tabular}
  \begin{minipage}{0.95\linewidth}
    \footnotesize{Note:} Results are reported as mean $\pm$ standard deviation across 100 independent results.
  \end{minipage}
  \vspace{-15pt}
\end{table*}

\begin{figure}[t]
    \centering
    \includegraphics[width=0.5\textwidth, keepaspectratio]{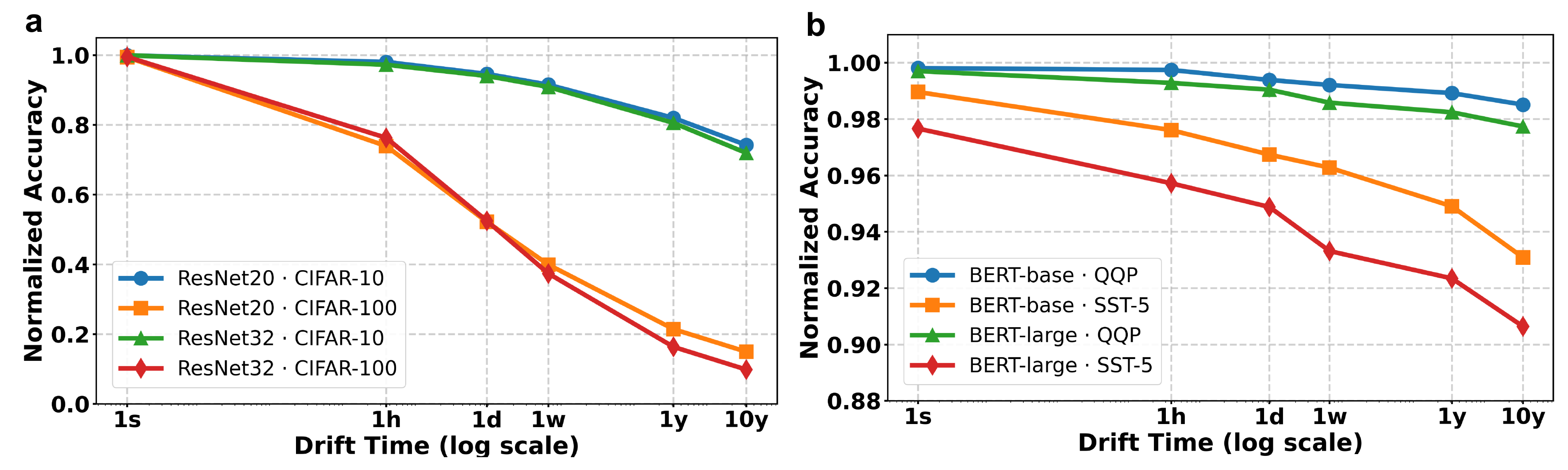}
    \vspace{-20pt}
    \caption{
Normalized accuracy degradation under drift for
    (a) CNNs and 
    (b) Transformers. 
    }
    
    \label{normal}
    \vspace{-12pt} 
\end{figure}

\begin{figure}[t]
    \centering
    \includegraphics[width=0.5\textwidth, keepaspectratio]{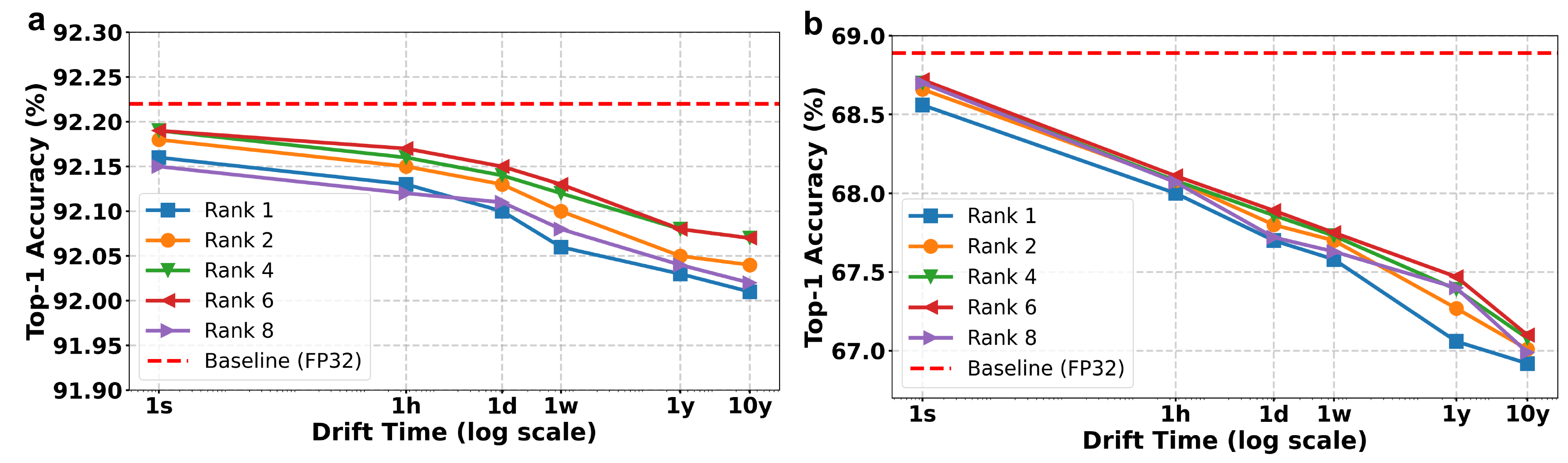}
    \vspace{-20pt}
    \caption{
Impact of low-rank dimension $r$ on VeRA+ compensation for ResNet-20. 
    (a) CIFAR-10. (b) CIFAR-100. 
    }\vspace{-4pt}
    \label{rank}
    \vspace{-6pt} 
\end{figure}

\section{Experimental Results}

\subsection{Experimental Setup}

We evaluate VeRA+ on both vision and language tasks. 
For image classification, we adopt ResNet-20/32 on CIFAR-10/100, ResNet50 on ImageNet-1k.
For NLP, we test BERT-base and BERT-large on QQP (paraphrase detection) and SST-5 (sentiment analysis). 
All models are quantized (W4A4 for ResNet, W4A8 for BERT) and evaluated under simulated conductance drift ranging from 1 second to 10 years.
We report normalized accuracy, defined as the ratio between the drifted or compensated model accuracy and its drift-free accuracy.

\subsection{Accuracy under Conductance Drift}

We first analyze accuracy degradation without compensation. 
Fig.~\ref{normal} shows the normalized accuracy drop across different tasks, and Table~\ref{tab:drift_results} summarizes representative results. 
Three key observations are made:  
(i) harder datasets (CIFAR-100, SST-5) degrade much faster than simpler ones (CIFAR-10, QQP), indicating that task complexity amplifies drift effects;  
(ii) CNNs (ResNet) are more vulnerable than Transformers, with ResNet models losing over 80\% accuracy on CIFAR-100 after 10 years, whereas BERT largely maintains performance, demonstrating stronger structural robustness.
(iii) among all evaluated models, ResNet-50 on ImageNet-1K suffers the most severe degradation due to the high complexity of the dataset, but VeRA+ still delivers a clear and substantial recovery in accuracy.

\subsection{VeRA+ Compensation Results with Various Rank $r$}

Table~\ref{tab:drift_results} demonstrates that VeRA+ effectively restores accuracy under severe long-term drift. 
For example, on ResNet-20 with CIFAR-100, the drift-free accuracy of 68.8\% drops to only 9.1\% after 10 years, 
while VeRA+ with $r=1$ recovers it to 66.9\%, restoring nearly 60 percentage points. 
Similar recovery is observed across all CNN benchmarks, and for language tasks, VeRA+ maintains almost drift-free accuracy 
(e.g., 90.05\% on QQP and 53.48\% on SST-5 for BERT-base and BERT-large), 
confirming its robustness across architectures.

Since RRAM-IMC systems primarily target CNN workloads~\cite{RIMC1,RIMC2,RIMC3}, and our earlier results indicate that BERT models already possess strong intrinsic drift tolerance, the following analysis focuses on CNNs.  

We further conduct an ablation study on the effect of rank $r$ in VeRA+. 
As shown in Fig.~\ref{rank}, experiments on ResNet-20 with CIFAR-10 (a) and CIFAR-100 (b) are performed under drift durations from 1 second to 10 years, with $r$ varied from 1 to 8.  
Even at $r=1$, VeRA+ significantly restores accuracy under severe drift (e.g., over 66.9\% on CIFAR-100 after 10 years), demonstrating that low-rank adaptation is highly effective. 
Accuracy improves steadily as $r$ increases up to 6, reflecting enhanced representational capacity; however, performance slightly declines at $r=8$, suggesting diminishing returns and possible overfitting. 
Similar trends have been observed in LLM fine-tuning~\cite{dora}, where overly large adaptation ranks reduce generalization. 
Considering both accuracy and efficiency, $r=1$ offers the best trade-off, achieving near-optimal robustness with minimal parameter and computation overhead.

\begin{figure}[t]
    \centering
    \includegraphics[width=0.5\textwidth, keepaspectratio]{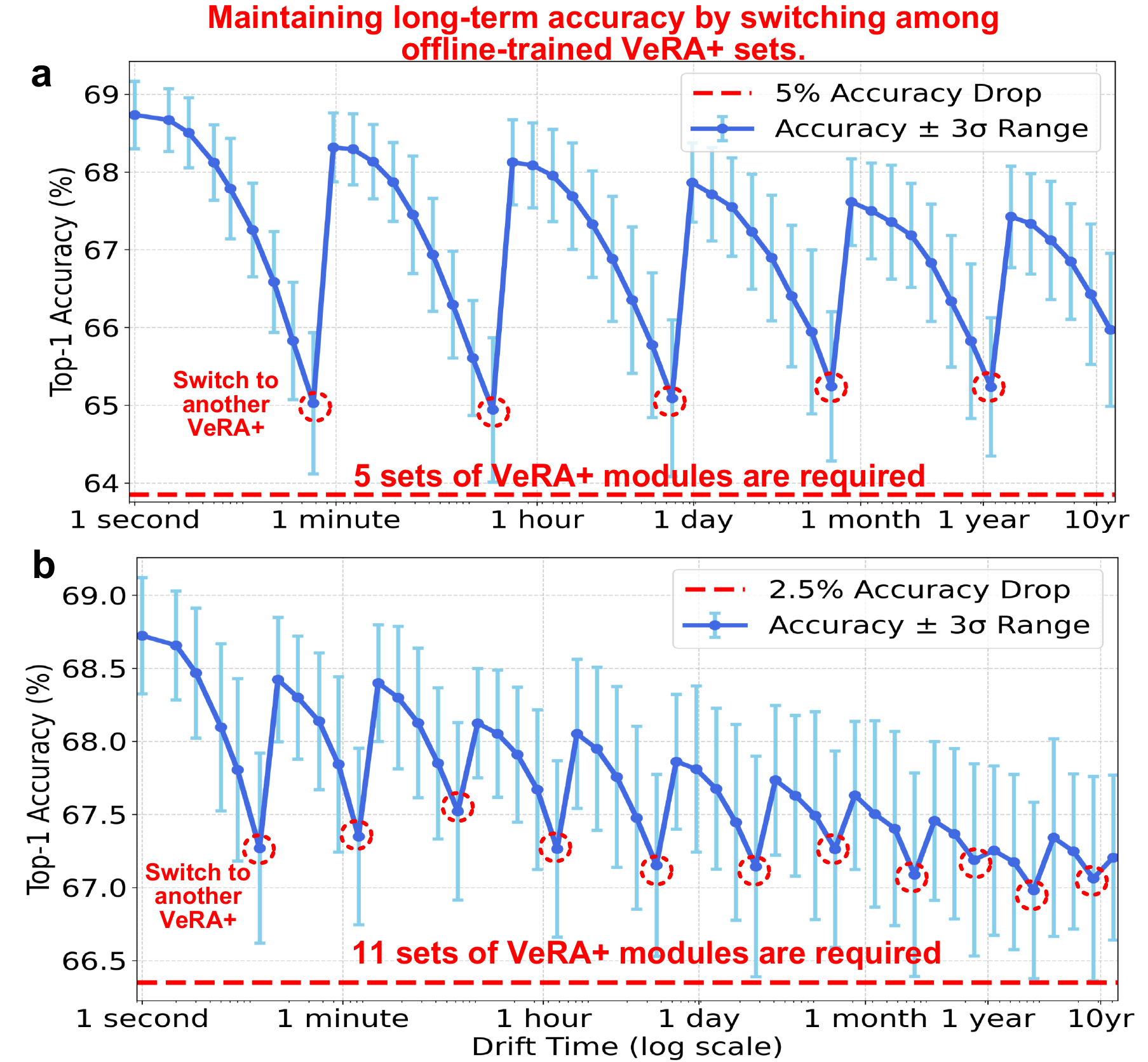}
    \vspace{-20pt}
    \caption{Number of required VeRA+ sets (r=1) under different acceptable accuracy drop thresholds. 
  }\vspace{-10pt}
    \label{fig_branch}
    \vspace{-16pt} 
\end{figure}

\begin{table}[t]
\centering
\caption{Parameter and operation overhead at $r=1$ with 11 sets for ResNet-20}
\vspace{-10pt}
\label{tab:overhead_paramops}
\begin{tabular}{lcc}
\toprule
\textbf{Method} & \textbf{Params Overhead} & \textbf{Ops Overhead} \\
\midrule
LoRA  & 47.0\% & 11.5\% \\
VeRA  & 11.9\% & 12.5\% \\
VeRA+ & 3.5\%  & 1.9\%  \\
\bottomrule
\vspace{-18pt}
\end{tabular}
\end{table}

\begin{table*}[t]
\centering
\small
\caption{Hardware Resource Comparison of Different Configurations on ResNet-20 with 11 Sets}
\vspace{-4pt}
\label{tab:hardware_results}
\begin{tabular}{ccccccccc}
\toprule
\textbf{Configuration} &
\makecell{\textbf{Area}\\(mm$^2$)} &
\makecell{\textbf{Area}\\Overhead} &
\makecell{\textbf{Energy}\\(nJ)} &
\makecell{\textbf{Energy}\\Overhead} &
\makecell{\textbf{Weight Data}\\Movement (KB)} &
\makecell{\textbf{Storage}\\(KB)} &
\makecell{\textbf{10y Normalized}\\Accuracy (CIFAR-10)} &
\makecell{\textbf{10y Normalized}\\Accuracy (CIFAR-100)} \\
\midrule
\rowcolor{red!10}Pure RRAM        & 0.429 & 0     & 210.2 & 0     & 0     & 0     & 53.94\% & 13.21\% \\
\rowcolor{green!10} VeRA+ rank = 1 & 0.444 & 3.5\% & 219.6 & 4.5\% & 2.40  & 5.15  & 99.77\% & 97.14\% \\
VeRA+ rank = 6   & 0.464 & 8.2\%  & 250.9 & 19.4\% & 5.56  & 6.45  & 99.83\% & 97.40\% \\
VeRA rank = 1    & 0.463 & 8.1\%  & 267.6 & 27.3\%  & 5.54  & 16.50  & 99.79\% & 97.17\% \\
VeRA rank = 6    & 0.631 & 47.1\% & 561.0 & 166.9\% & 31.93 & 30.15  & 99.84\% & 97.44\% \\
LoRA rank = 1    & 0.582 & 35.6\% & 266.8 & 26.9\%  & 24.21 & 66.52  & 99.82\% & 97.33\% \\
 LoRA rank = 6 & 1.349 & 214.4\% & 556.4 & 164.7\% & 145.96 & 401.18 & 99.85\% & 97.64\% \\
\bottomrule
\vspace{-15pt}
\end{tabular}
\end{table*}



\subsection{Accuracy–Efficiency Trade-off}

Fig.~\ref{fig_branch} shows the relationship between the accuracy tolerance and the number of VeRA+ sets required for reliable inference. 
As expected, tighter accuracy constraints demand finer-grained compensation. 
 
For example, allowing a 5\% accuracy drop requires 5 pre-trained sets, while tightening the tolerance to 2.5\% increases the count to 11. 
This result highlights a clear trade-off in VeRA+ scheduling: stricter accuracy preservation improves robustness to worst-case drift but incurs more sets of compensation modules. For each drift time $t$, 100 independent drift instances are simulated to estimate the mean accuracy $\mu$ and its standard deviation $\sigma$.

\subsection{Hardware Resources and Accuracy}
Table~\ref{tab:overhead_paramops} summarizes the parameter and
operation overhead of different adaptation methods at $r=1$ with
11 drift sets. VeRA+ introduces only 3.5\% parameter and 1.9\%
operation overhead, over 80\% lower than LoRA, demonstrating
that it remains highly lightweight even when supporting many drift
levels.

We estimate the hardware cost of drift-aware compensation under a 22\,nm CMOS process. 
The total energy is approximated as
\begin{equation}
E = \frac{\mathrm{Ops}_{\mathrm{RRAM}}}{\mathrm{TOPS/W}_{\mathrm{RRAM}}}
  + \frac{\mathrm{Ops}_{\mathrm{SRAM}}}{\mathrm{TOPS/W}_{\mathrm{SRAM}}},
\label{eq:energy_estimate}
\end{equation}
where $\text{TOPS/W}$ values are taken from Table~\ref{tab:hardware}. 
We assume the SRAM-IMC can hold one set of compensation parameters, giving a conservative upper bound on area, though streaming and tiling can further reduce the SRAM-IMC area requirement~\cite{wj}.

Table~\ref{tab:hardware_results} summarizes the estimates. 
The pure RRAM baseline offers the smallest area but degrades sharply after 10 years (53.94\% on CIFAR-10, 13.21\% on CIFAR-100).  
With VeRA+ ($r=1$), accuracy recovers to 99.77\% and 97.14\% with only 3.5\% area and 4.5\% energy overhead; $r=6$ improves slightly to 99.83\% and 97.40\% at 8.2\% area and 19.4\% energy.  
In contrast, VeRA requires 8.1–47.1\% area and 27.3–166.9\% energy, while LoRA is far more expensive (35.6–214.4\% area, 26.9–164.7\% energy). Overall, VeRA+ provides near drift-free accuracy with over 5× lower overhead than VeRA and 10× lower than LoRA, achieving the best lightweight compensation for long-term RRAM-IMC deployment.

\subsection{Comparison with BN-based Calibration}

Table~\ref{tab:bn_compare} compares VeRA+ with BN-based calibration ~\cite{variationaware} as a SOTA method.  Both methods recover accuracy close to the drift-free baseline on ResNet-20
(CIFAR-10). However, BN-based methods require 5\% of the training set and performing
online BN-statistics recomputation, leading to 7.5\,MB storage and 1.8\%
operation overhead for ResNet-20 on CIFAR-10.  
VeRA+ avoids data replay and online calibration entirely and requires only
5.15\,KB of parameters with comparable operation cost (1.9\%), offering over
a 1000$\times$ reduction in storage overhead.

\begin{table}[t]
\centering
\vspace{-8pt}
\caption{Comparison between BN-based calibration  and VeRA+ for ResNet-20 (CIFAR-10).}
\label{tab:bn_compare}
\vspace{-8pt}
\begin{tabular}{lccc}
\toprule
\textbf{Method} &
\makecell{\textbf{Storage}\\\textbf{Overhead}} &
\makecell{\textbf{Operation}\\\textbf{Overhead}} &
\makecell{\textbf{On Chip}\\\textbf{Calibration}} \\
\midrule
BN-based \cite{variationaware} & 7.5\,MB  & \cellcolor{green!10}1.8\%  & Yes \\
VeRA+           & \cellcolor{green!10}5.15\,KB & 1.9\%  & \cellcolor{green!10}No \\
\bottomrule
\end{tabular}
\vspace{-8pt}
\end{table}

\begin{figure}[t]
    \centering
    
    \includegraphics[width=0.5\textwidth, keepaspectratio]{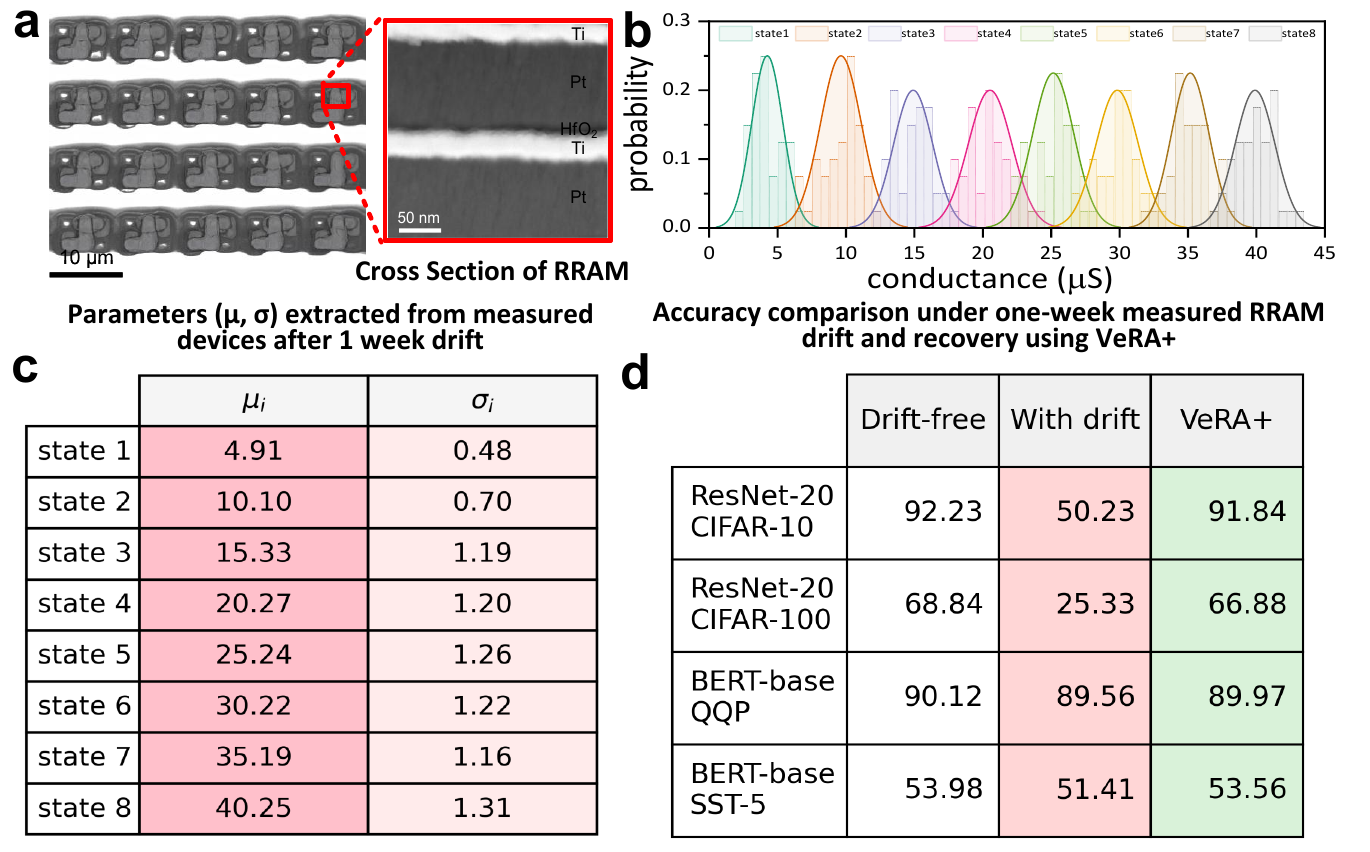}
    \vspace{-15pt} 
    \caption{(a) RRAM device structure, (b) conductance statistics, (c) drift parameters $(\mu_i, \sigma_i)$ of one week drift  (d) and VeRA+ drift compensation results. 
  } 
    \vspace{-10pt}
    \label{fig:rram_data}
    \vspace{-10pt} 

\end{figure}

\subsection{Validation under Realistic Device Drift}

 To validate VeRA+ under realistic device behaviors, we first characterized a Ti/HfO\textsubscript{x}/Pt 1T1R RRAM array fabricated in a 180-nm CMOS process. The RRAM stack was integrated using BEOL processing on top of the peripheral circuits, as shown in Fig.~\ref{fig:rram_data}. Eight conductance levels (5–40 µS) were programmed by tuning the access-transistor compliance current and measured at a 0.2 V read voltage. All measurements were collected one week after programming to capture realistic temporal drift. For each programmed state, 200 devices were measured to extract a state-dependent Gaussian drift model, yielding per-state $(\mu_i, \sigma_i)$ values. These state-dependent drift parameters were then used to replace the IBM drift model in Section~\ref{subsec:driftmodel} for training the VeRA+ compensation vectors.

We also mapped the full ResNet-20 weight matrix onto five 256×512 RRAM arrays. One week after programming, the full conductance map was read out and converted back into network weights, which were evaluated on CIFAR-10 and CIFAR-100 to quantify real drift-induced degradation. We then applied VeRA+ compensation using the same hardware-extracted drift parameters and re-evaluated the accuracy. For the BERT models, whose parameters are too large to fit on our measured arrays, we instead sampled drift noise directly from the measured drift model and performed software-based drift simulations. Across both the one-week hardware measurements and the measured-drift simulations, VeRA+ restores accuracy close to the drift-free baseline (Fig.~\ref{fig:rram_data}(d)), demonstrating strong robustness to real, non-uniform RRAM drift and confirming its practicality under realistic hardware conditions.

\section{Conclusion}
We presented VeRA+, a lightweight drift-compensation framework
for RRAM-based in-memory computing. By reusing shared
projection matrices and introducing only two compact
drift-specific scaling vectors, VeRA+ achieves efficient, scalable,
and hardware-friendly adaptation across both CNNs and
Transformers. Compared with LoRA-style adaptation, VeRA+ delivers similar compensation capability at a fraction of the parameter and compute overhead   Validation on 180-nm 1T1R RRAM devices further confirms its
robustness under realistic, state-dependent drift behavior. The proposed approach is
 potentially applicable to other non-volatile IMC platforms such
 as PCM, MRAM, and Flash, though platform-specific drift
 characterization and experimental validation remain necessary.

\end{document}